\documentclass[a4paper,reqno,12pt]{amsart}
\usepackage{amssymb,bbold}
\usepackage{array}
\usepackage{fancybox}
\newcommand{\repre}[1]{\underline{\mathbf{#1}}}
\newcommand{\so}{\mathfrak{so}}
\newcommand{\SO}{\mathrm{SO}}
\newcommand{\GL}{\mathrm{GL}}
\newcommand{\U}{\mathrm{U}}
\newcommand{\SU}{\mathrm{SU}}
\newcommand{\CC}{\mathbb{C}}
\newcommand{\M}{\mathsf{M}}
\newcommand{\RR}{\mathbb{R}}
\newcommand{\EE}{\mathbb{E}}
\newcommand{\HH}{\mathbb{H}}
\newcommand{\Spin}{\mathrm{Spin}}
\newcommand{\spin}{\mathfrak{spin}}
\newcommand{\Cl}{\mathrm{C}\ell}
\newcommand{\1}{\mathbb{1}}
\newcommand{\ii}{\mathbb{i}}
\newcommand{\jj}{\mathbb{j}}
\newcommand{\kk}{\mathbb{k}}
\newcommand{\half}{\tfrac{1}{2}}
\newcommand{\ZZ}{\mathbb{Z}}
\newcommand{\OO}{\mathbb{O}}
\newcommand{\fg}{\mathfrak{g}}
\newcommand{\su}{\mathfrak{su}}
\renewcommand{\u}{\mathfrak{u}}
\renewcommand{\Re}{\mathrm{Re}\,}
\renewcommand{\Im}{\mathrm{Im}\,}
\renewcommand{\Sp}{\mathrm{Sp}}
\renewcommand{\sp}{\mathrm{sp}}
\renewcommand{\O}{\mathrm{O}}
\DeclareMathOperator{\vol}{\mathrm{vol}}
\DeclareMathOperator{\Mat}{Mat}
\DeclareMathOperator{\tr}{Tr}
\DeclareMathOperator{\slg}{SLAG}
\DeclareMathOperator{\clg}{\mathbb{C}LAG}
\begin{document}

\title{Intersecting brane geometries}
\author{Jos\'e M Figueroa-O'Farrill}
\address{\begin{flushright}
Department of Physics\\
Queen Mary and Westfield College\\
Mile End Road\\
London E1 4NS, UK\end{flushright}}
\email{j.m.figueroa@qmw.ac.uk}
\date{\today}
\begin{abstract}
We present a survey of the calibrated geometries arising in the study
of the local singularity structure of supersymmetric fivebranes in
$\M$-theory.  We pay particular attention to the geometries of
$4$-planes in eight dimensions, for which we present some new results
as well as many details of the computations.  We also analyse the
possible generalised self-dualities which these geometries can
afford.
\end{abstract}
\maketitle

\tableofcontents

\section{Introduction}

Recent developments suggest that the natural language in which to
phrase the study of intersecting branes is that of calibrated
geometry.  This has proven effective both in the description of their
singularity structure
\cite{GibbonsPapadopoulos,AFS-cali,AFS-groups,AFSS} as in the study of
the geometry of branes within branes \cite{GLW}.  In our recent work
\cite{AFS-cali,AFS-groups,AFSS} on the singularity structure of
intersecting branes, which continues the approach laid out in
\cite{Townsend-Mbranes,OhtaTownsend}, we encountered a number of
calibrated geometries, not all of which are well-known.  We believe
that it could be useful to present a detailed survey of these
geometries in the present context.  This is one of the purposes of the
present paper.  Manifolds admitting special geometries of the kind
described here also admit generalisations of the notion of
self-duality.  In the context of gauge theory, these generalised
self-duality give rise to higher-dimensional generalisations of the
notion of instanton.  We think it useful to also work out the possible
generalisations of self-duality that these geometries give rise to.

This paper is organised as follows. We start in Section
\ref{sec:whatis} by explaining what is meant by a {\em geometry\/} in
this context and how to translate between its local and global (for
lack of a better name) descriptions.  Then in Section
\ref{sec:results} we will explain how a configuration of intersecting
branes gives rise to a geometry, and present a list of all the known
geometries which arise in this way.  Section \ref{sec:geometries} is
devoted to a case-by-case description of each of these geometries.
Section \ref{sec:tour} presents a detailed ``guided tour'' of the
different eight-dimensional geometries.  The calculational details
presented in this section might be of some help to people working in
this topic.  In Section \ref{sec:selfdual} we explore the different
generalisations of self-duality which these calibrated geometries
afford.  For some of the less common geometries, these results are
new.  Finally in Section \ref{sec:fin} we make some concluding
remarks.

\section{What is meant by a geometry?}
\label{sec:whatis}

Generally speaking, a geometry is some sort of structure we endow a
manifold with.  Traditionally a geometry is specified through the
existence of certain tensor fields on the manifold.  Many well-known
geometries arise in this way. For example, a metric gives rise to
riemannian geometry, a closed nondegenerate $2$-form to symplectic
geometry, and a complex structure gives rise to complex geometry.  We
can add more structure to one or more of these and obtain many of the
geometries which have taken centre stage in recent times: K\"ahler,
hyperk\"ahler, quaternionic K\"ahler,...  This way of defining a
geometry often goes hand-in-hand with a reduction of the structure
group of the frame bundle.  On a differential $n$-manifold without
extra structure, the frame bundle is a principal $\GL_n\RR$ bundle.
With the introduction of a metric, we can consistently restrict
ourselves to orthonormal frames, and in effect reduce the structure
group to $\O_n$.  Similarly, if $n=2m$, a complex structure allows us
to consider complex frames so that the structure group reduces to
$\GL_m\CC$.  More generally, a $G$-structure on a manifold is a
reduction of structure group of the frame bundle to
$G\subset\GL_n\RR$, so that one can choose local frames which are
$G$-related.  Not every manifold admits any $G$-structure: there might
be topological obstructions.  For example, although every manifold
admits a metric and hence an $\O_n$ structure, unless the manifold is
orientable it will not admit an $\SO_n$ structure.

A common way to reduce the structure group to $G\subset\O_n$ is via a
metric whose holonomy lies in $G$; although not all $G\subset\O_n$ can
be so realised, unless one simultaneously allows for torsion in the
metric connection.  Given a riemannian manifold $M$ whose metric has
holonomy $G$, the holonomy principle \cite{Besse} guarantees the
existence of privileged tensors on $M$ corresponding to those
(algebraic) tensors which are $G$-invariant.  For example, this lies
at the heart of the equivalence between the two common definitions of
a K\"ahler manifold: as a riemannian $2m$-di\-men\-sion\-al manifold
with $\U_m$ holonomy, or as a riemannian manifold with a parallel
complex structure---the complex structure and the metric both being
$\U_m$-invariant tensors.  This way of specifying a geometry has
played an important role in superstring theory, via the study of
supersymmetric sigma models particularly.

More recently, however, with the advent of branes, another way of
specifying a geometry has become increasingly relevant.  Instead of
using the existence of tensorial objects or of reductions of the
structure group, one specifies a geometry on a manifold by singling
out a special class of submanifolds.  For example, one can talk about
minimal submanifolds of a riemannian manifold, or about complex
submanifolds of a complex manifold.  In fact, in a K\"ahler manifold
these two kinds of submanifolds are not unrelated: complex
submanifolds are minimal.  The theory of calibrations provides a
systematic approach to understanding this fact and allows, in
addition, for far-reaching generalisations of this statement.
In order to facilitate the discussion it will be necessary to first
introduce a few important concepts.

\subsection{Submanifolds and grassmannians}

Let $N$ be a $p$-di\-men\-sion\-al submanifold of an
$n$-di\-men\-sion\-al manifold $M$.  At each point $x\in N$, the
tangent space $T_xN$ to $N$ is a $p$-di\-men\-sion\-al subspace of the
tangent space $T_xM$.  In a small enough neighbourhood $U$ of $x$, we
can trivialise the tangent bundle of $M$.  This means essentially that
we can identify the tangent space to any point in $U$ with $\RR^n$.
Now consider those points $y$ in $U$ which also lie in $N$.  Under
this identification, the tangent space $T_yN$ of $N$ at $y$ will be
identified with a $p$-plane in $\RR^n$.  This defines a map from
$N\cap U$ to the space of $p$-planes in $\RR^n$.  Spaces of planes are
generically called {\em grassmannians\/} and will play a central role
in the following discussion, so it pays to take a brief look at them
before going further.

It is convenient to identify $p$-planes with certain types of
$p$-vectors.  The identification runs as follows.  Let $\pi$ be a
$p$-plane in $\RR^n$.  Let $e_1,e_2,\ldots,e_p$ be a basis for $\pi$.
Then the $p$-vector $e_1\wedge e_2 \wedge \cdots \wedge e_p$ is
nonzero.  However, if we choose a different basis
$e'_1,e'_2,\ldots,e'_p$ for $\pi$, then we generally end up with a
different $p$-vector $e'_1\wedge e'_2 \wedge \cdots \wedge e'_p$.  Of
course, both $p$-vectors are proportional to each other, the constant
of proportionality being the (nonzero) determinant of the linear
transformation which takes one basis to the other.  Conversely, given
a non-zero $p$-vector $v_1\wedge v_2 \wedge \cdots \wedge v_p$, we
associate with it the $p$-plane $\pi$ spanned by the $\{v_i\}$, with
the proviso that as above, proportional $p$-vectors give rise to the
same $p$-plane.  We can eliminate the multiplicative ambiguity by
picking a privileged $p$-vector for each plane.  This can be done by
introducing a metric in $\RR^n$ and considering only oriented planes.
We will reflect this fact by saying that we consider oriented
$p$-planes in the euclidean space $\EE^n$.

Let $G(p|n)$ denote the grassmannian of oriented $p$-planes in
$\EE^n$.  As we now show it can be identified with a subspace of the
unit sphere in $\EE^{\binom{n}{p}}$.  Indeed, given an oriented
$p$-plane $\pi$, let $e_1,e_2,\ldots,e_p$ be an oriented orthonormal
basis and consider the $p$-vector $e_1 \wedge e_2 \wedge \cdots \wedge
e_p \in \bigwedge^p \EE^n$, which we will also denote $\pi$
consistently with the identification we are describing.  The norm of
any $p$-vector $v_1 
\wedge v_2 \wedge \cdots \wedge v_p$ is given by
\begin{equation*}
\| v_1 \wedge v_2 \wedge \cdots \wedge v_p \| = | \det\langle
v_i,v_j\rangle|~.
\end{equation*}
This norm extends to a metric on $\bigwedge^p \EE^n$, which turns it
into a euclidean space $\EE^{\binom{n}{p}}$.  It follows that $\pi$ has
unit norm, so that it belongs to the unit sphere in
$\EE^{\binom{n}{p}}$.  Conversely, every simple (i.e., decomposable
into a wedge product of $p$ vectors) unit $p$-vector $\pi = e_1 \wedge
e_2 \wedge \cdots \wedge e_p \in \bigwedge^p \EE^n$ defines an
oriented $p$-plane with basis $e_1,e_2,\ldots,e_p$.  In other words,
$G(p|n)$ can be identified with a subset of the unit sphere in
$\EE^{\binom{n}{p}}$, so that it is a compact space.  This can also be
understood from the fact that the grassmannian $G(p|n)$ is acted on
transitively by $\SO_n$.  Indeed it is not hard to see that the
isotropy consists of changes of basis in $\pi$ and its perpendicular
$(n-p)$-plane $\pi^\perp$; whence it is isomorphic to
$\SO_p\times\SO_{n-p}$.  This means that the grassmannian is a coset
manifold:
\begin{equation*}
G(p|n) \cong \frac{\SO_n}{\SO_p\times\SO_{n-p}}\cong G(n-p|n)~.
\end{equation*}

\subsection{Geometries and grassmannians}

As mentioned above, a geometry can be specified by singling out a
class of special submanifolds.  For example, one could consider
submanifolds whose tangent spaces belong to a certain subset of the
grassmannian of planes.  These give rise to so-called {\em
grassmannian geometries\/}.  A special type of subset of the
grassmannian $G(p|n)$ are those sets which correspond to the orbit of
a plane under a subgroup of $\SO_n$.  It will turn out that all the
geometries that we will encounter will be of this form.

For example, suppose that $n=2m$.  Then we could consider complex
$k$-dimensional submanifolds; that is, $p=2k$.  The tangent subspaces
to these submanifolds are $k$-dimensional complex subspaces of
$\CC^m\cong\RR^n$.  All the tangent planes belong to the $\U_m \subset
\SO_n$ orbit of any one of the planes.  The resulting orbit is the
complex grassmannian $G_\CC(k|m)$ of $k$-dimensional complex planes in
$\CC^m$.  It is not hard to see that
\begin{equation*}
G_\CC(k|m) \cong \frac{\U_m}{\U_k\times\U_{m-k}} \cong
\frac{\SU_m}{\mathrm{S}\left(\U_k\times\U_{m-k}\right)}~,
\end{equation*}
so that, in fact, the planes belong to the same $\SU_m$ orbit.

Similarly if $n=4\ell$, we can consider quaternionic subspaces of
$\HH^\ell\cong\RR^n$.  They necessarily have dimension $p=4j$.  The
grassmannian $G_\HH(j|\ell)$ of quaternionic planes correspond to the
orbit of a plane under $\Sp_\ell \subset \SO_n$, so that
\begin{equation*}
G_\HH(j|\ell)\cong \frac{\Sp_\ell}{\Sp_j\times \Sp_{\ell-j}}~.
\end{equation*}

Other examples are possible, and we shall discuss them below.  For now
let us simply point out the fact that for the complex and quaternionic
grassmannians, the subgroups $\SU_m$ and $\Sp_\ell$ of $\SO_n$ are
such that they (or their lifts to subgroups of $\Spin_n$) leave some
spinors invariant.  This is intimately linked to supersymmetry and
will also be the case for the other examples we will encounter.  We
now turn our attention to another way to single out subsets of the
grassmannian.

\subsection{Calibrations}

Calibrations will provide us a tool with which to specify subsets
(faces, actually) of the grassmannian of planes.  The geometries which
are obtained in this fashion are known as {\em calibrated
geometries\/}.  The foundations of this subject are clearly explained
in \cite{HarveyLawson}, and a shorter but lucid exposition can be
found in \cite{Morgan-monthly}.

Let $\varphi \in \bigwedge^p \left(\EE^n\right)^*$ be a (constant
coefficient) $p$-form on $\EE^n$.  It defines a linear function on
$\bigwedge^p \EE^n$, which restricts to a continuous function on the
grassmannian $G(p|n)$.  Because $G(p|n)$ is compact, this function
attains a maximum, called the {\em comass\/} of $\varphi$ and denoted
$\|\varphi\|^*$.  If $\varphi$ is normalised so that it has comass 1,
then it is called a {\em calibration\/}.  Let $G(\varphi)$ denote
those points in $G(p|n)$ on which $\varphi$ attains its maximum.
$G(\varphi)$ is known as the {\em $\varphi$-grassmannian\/}, and
planes $\pi \in G(\varphi)$ are said to be {\em calibrated by\/}
$\varphi$.  The subset $\cup_\varphi G(\varphi) \subset G(p|n)$, where
the union runs over all calibrations $\varphi$, defines the {\em faces
of $G(p|n)$\/}.  The name comes from the fact that if we think of
$G(p|n)$ as a subset of the vector space $\EE^{\binom{n}{p}}$, then
$G(\varphi)$ is the contact set of $G(p|n)$ with the hyperplane
$\{\xi\in\EE^{\binom{n}{p}}\mid\varphi(\xi)=1\}$.  Now, because
$\varphi$ is a calibration, $\varphi(\xi) \leq 1$ and hence $G(p|n)$
lies to one side of that hyperplane.

Computing the comass of a $p$-form is a difficult problem which has
not been solved but for the simplest of forms $\varphi$, those which
have a high degree of symmetry or those which can be obtained by
squaring spinors.  Determining the faces of the grassmannian has
proven equally difficult and has only been achieved completely in the
lowest dimensions.  The determination of the faces of the grassmannian
$G(p|n)$ is not an easy problem whenever $p$ is different from $1$,
$2$, $n-2$, or $n-1$.  To this day, only the cases $n = 6$
\cite{DadokHarvey-R6,HarveyMorgan-6,Morgan-ext} and $n=7$
\cite{HarveyMorgan-7,Morgan-ext} have been fully solved, whereas there
are some partial results for $n=8$ \cite{DadokHarveyMorgan}.  In the
study of static fivebranes in $\M$-theory it is the case $n=10$ that
is needed.

A $p$-submanifold $N$ of $\EE^n$, all of whose tangent
planes belong to $G(\varphi)$ for a fixed calibration $\varphi$, is
said to be a {\em calibrated submanifold\/}.  A calibrated submanifold
$N$ has minimum volume among the set of all submanifolds $N'$ with the
same boundary.  This is because
\begin{equation*}
\vol N = \int_N \varphi = \int_{N'} \varphi \leq \vol N'~,
\end{equation*}
where the second equality follows by Stokes' theorem.  Calibrated
submanifolds constitute a far-reaching generalisation of the notion of
a geodesic.  Indeed, the grassmannian of oriented lines $G(1|n)$ is
just the unit sphere $S^{n-1}\subset\EE^n$, whose faces are obviously
points.  Hence the tangent spaces of a one-dimensional submanifold $L$
belong to the same face if and only if $L$ is a straight line.  Notice
that there is a duality between $p$-dimensional and $p$-codimensional
submanifolds; in fact, if $\varphi$ is a calibration so is
$\star\varphi$.  Hence hyperplanes in $\EE^n$ are also (locally)
volume-minimising.

This theory is not restricted to constant coefficient calibrations in
$\EE^n$.  In fact, we can work with $d$-closed forms $\varphi$ in any
riemannian manifold $(M,g)$.  The comass of $\varphi$ is now the
supremum (over the points in $M$) of the comasses at each point.  If
$M$ is compact, this supremum exists.  A calibration is now a
$d$-closed form normalised to have unit comass; or equivalent one
which satisfies
\begin{equation*}
\varphi_x(\xi) \leq \vol \xi \qquad\text{for all oriented tangent
$p$-planes $\xi$ at $x$.}
\end{equation*}
Notice that there may be points in $M$ for which the
$\varphi$-grassmannian is empty.  The same argument as before shows
that calibrated submanifolds are homologically volume-minimising.
Of course, this crucially necessitates that $\varphi$ be $d$-closed.

If an oriented riemannian $n$-manifold has reduced holonomy, meaning a
proper subgroup $G$ of $\SO_n$, then the holonomy principle guarantees
the existence of parallel (hence $d$-closed) forms corresponding to
the $G$-invariants in the exterior power of the tangent
representation.  It turns out that in many (if not all) cases, the
parallel forms are calibrations giving rise to interesting geometries.
The $\varphi$-grassmannian associated to a $G$-invariant form
$\varphi$ contains, and in many cases coincides with, the $G$-orbit of
any one of its planes.

\section{The (local) geometry of intersecting branes}
\label{sec:results}

In this section we summarise the results of
\cite{AFS-cali,AFS-groups,AFSS} and tabulate the different geometries
that were found.  These geometries will be described in more detail in
Section \ref{sec:geometries}.

\subsection{From branes to geometry}

Branes can be understood as certain types of solutions to the
supergravity equations of motion.  These solutions are characterised
by their invariance (at least locally) under a
$(p+1)$-di\-men\-sion\-al super-Poincar\'e subalgebra.  The solutions
describe the exterior spacetime to the worldvolume of a
$p$-dimensional extended object: the brane.  The brane therefore
corresponds to a $(p+1)$-Lorentzian submanifold, with possible
self-intersections.  In many cases these submanifolds are minimal and
just as for minimal immersions
\cite{Morgan-angle,Nance,Lawlor,Lawlor-kplanes} one can ask what is
the local singularity structure of a brane solution.

For definiteness we will only discuss fivebranes in
eleven-di\-men\-sion\-al supergravity in this note.  It is clear
that this approach generalises to general $p$-branes in this and other
supergravities; although it may be possible to treat more general
cases from this one by using duality transformations.

Let $B$ be the worldvolume of a fivebrane in an eleven-dimensional
spin manifold $M$.  Fix a point $x\in B$.  Choosing an orthonormal
frame\footnote{Following \cite{Townsend-Mbranes} we employ the symbol
$\natural$ (pronounced `ten') to refer to the tenth spatial
coordinate.} $e_0,e_1,\ldots,e_9,e_\natural$ for the tangent space
$T_xM$ to $M$ at $x$, we can identify $T_xM$ with
eleven-di\-men\-sion\-al Minkowski spacetime $\EE^{10,1}$.  The 
tangent spaces (if $x$ is a singular point of the immersion then there
is more than one) to the worldvolume of a fivebrane passing through
$x$ define a subset of the grassmannian $G(5,1|10,1)$ of oriented
time-oriented $(5,1)$-planes in $\EE^{10,1}$, which analogously to the
euclidean case, is a coset space
\begin{equation*}
G(5,1|10,1) \cong
\frac{\SO^\uparrow_{10,1}}{\SO^\uparrow_{5,1}\times\SO_5}~,
\end{equation*}
where $\SO^\uparrow$ stands for the connected component of the
identity.  The requirement of supersymmetry constraints which
subsets of this grassmannian can the tangents to the branes belong
to.

\subsection{Supersymmetry}

Let $\Cl_{1,10}$ be the Clifford algebra associated to $\EE^{10,1}$,
but with the opposite norm.  In other words, if $v\in\EE^{10,1}$ then
its Clifford square in $\Cl_{1,10}$ is given by
\begin{equation*}
v \cdot v = + \|v\|^2~\1~,
\end{equation*}
where $\|v\|^2 \equiv -(v^0)^2 + (v^1)^2 + \cdots + (v^\natural)^2$.
As associative algebras $\Cl_{1,10} \cong \Mat_{32}(\RR) \oplus
\Mat_{32}(\RR)$, whence it has two inequivalent irreducible
representations, each real and $32$-dimensional.  They are
distinguished by the action of the volume element, which takes the
values $\pm1$.  Fix one of these irreducible representations $\Delta$
once and for all---the choice is immaterial because they are both
equivalent under $\Spin_{10,1}\subset\Cl_{1,10}$.  Every $(5,1)$-plane
$\pi$ in $\EE^{10,1}$ defines a subspace
\begin{equation*}
\Delta(\pi) \equiv \{\psi\in\Delta \mid \pi \cdot \psi = \psi\}~,
\end{equation*}
where $\cdot$ stands for Clifford action and where we have used
implicitly the isomorphism of the Clifford algebra $\Cl_{1,10}$ with
the exterior algebra.  The subspace $\Delta(\pi)$ is non zero.  In
fact, because $\pi$ has unit norm, so that $\pi \cdot \pi = \1$, and
zero trace, $\Delta(\pi)\subset\Delta$ is $16$-dimensional.

If $\pi_1\equiv \pi,\pi_2,\ldots,\pi_m$ are $m$ $(5,1)$-planes, then
we say that the configuration $\cup_{i=1}^m \pi_i$ is {\em
supersymmetric\/} if and only if
\begin{equation*}
\Delta(\cup_{i=1}^m \pi_i) \equiv \bigcap_{i=1}^m \Delta(\pi_i) \neq
\{0\}~.
\end{equation*}
Moreover, such a supersymmetric configuration is said to preserve a
fraction $\nu$ of the supersymmetry, whenever
\begin{equation*}
32 \nu = \dim \Delta(\cup_{i=1}^m \pi_i)~.
\end{equation*}
A priori $\nu$ can only take the values $\tfrac{1}{32}, \tfrac{1}{16},
\tfrac{3}{32}, \ldots, \half$; although only the following fractions
are known to occur: $\tfrac{1}{32}$, $\tfrac{1}{16}$, $\tfrac{3}{32}$,
$\tfrac{1}{8}$, $\tfrac{5}{32}$, $\tfrac{3}{16}$, $\tfrac{1}{4}$ and
$\tfrac{1}{2}$.  From the full solution \cite{OhtaTownsend,AFSS} of
the two fivebrane problem it follows that there are no configurations
with fraction $\tfrac14 < \nu < \half$.  Therefore the only possible
fraction which has yet to appear is $\tfrac{7}{32}$.

A brane $B$ such that its tangents define a supersymmetric
configuration is called a {\em supersymmetric brane\/}.  An important
problem in this topic is the classification of the possible
supersymmetric configurations of so-called intersecting branes (see
\cite{Jerome} for a recent review and guide to the literature).  Each
such configuration gives rise to a subset of the grassmannian and, by
the discussion in Section \ref{sec:whatis}, to a geometry which, as we
will see, turns out to be calibrated.  This follows from the
correspondence between spinors and calibrations, to which we now turn.

\subsection{Calibrations and spinors}

The relationship between spinors and calibrations is well documented.
Although computing the comass of a form $\varphi$ is generally a
difficult problem, it simplifies tremendously when $\varphi$ can be
constructed by squaring spinors.  The cleaner results are in seven and
eight dimensions \cite{LM,Harvey} and more generally in $8k$
dimensions \cite{DadokHarvey-spinors}; but similar results can also be
obtained in eleven dimensions with Lorentzian signature
\cite{AFS-cali}.  Remarkably, it is the eleven- and eight-dimensional
cases which arise in the study of intersecting branes
\cite{AFS-cali,AFSS}.

\subsubsection*{Eight dimensions}

Let us first discuss the eight-dimensional case.  As an associative
algebra, the Clifford algebra $\Cl_8$ is isomorphic to the matrix
algebra $\Mat_{16}(\RR)$.  This means that it has a unique irreducible
representation $\Delta$ which is real and has dimension $16$.  Under
the spin group $\Spin_8 \subset \Cl_8$, $\Delta$ breaks up as
$\Delta_+ \oplus \Delta_-$, where each $\Delta_\pm$ corresponds to
spinors of definite chirality.  Let $\psi \in \Delta_+$ be a
chiral spinor, and consider the bispinor $\psi\otimes\bar\psi$.  It is
an element of $\Cl_8$ which, normalising the spinor appropriately, can
be written as
\begin{equation}\label{eq:squaring8d}
\psi \otimes \bar\psi = \1 + \Omega + \vol~,
\end{equation}
where $\Omega$ is a self-dual $4$-form in $\EE^8$.  Now let $\xi$ be a
simple unit $4$-vector in $\EE^8$.  Then it follows from the
expression of the bispinor that $\Omega(\xi) \|\psi\|^2 = \langle
\psi, \xi \cdot \psi\rangle$, where $\|\psi\|^2 = \langle\psi,
\psi\rangle$ is the norm relative to the natural $\Spin_8$-invariant
inner product on $\Delta_+$.  By the Cauchy--Schwarz inequality, it
follows that
\begin{equation*}
\Omega(\xi) = \frac{\langle \psi, \xi\cdot \psi\rangle}{\|\psi\|^2}
\leq \frac{\|\xi \cdot \psi\|}{\|\psi\|}~.
\end{equation*}
Because $\xi$ belongs to $\Spin_8 \subset \Cl_8$, $\|\xi \cdot
\psi\| = \|\psi\|$, whence $\Omega(\xi) \leq 1$ for all
$\xi$.  In other words, $\Omega$ has unit comass; that is, it is a
calibration.  It follows from this argument that the plane defined by
the 4-vector $\xi$ is calibrated by $\Omega$ if and only if $\xi
\cdot \psi = \psi$.

What can one say about the $\Omega$-grassmannian?  The isotropy of a
chiral spinor $\psi \in \Delta_+$ is a certain $\Spin_7^+$ subgroup of
$\Spin_8$, under which both $\Delta_-$ and the vector representation
of $\Spin_8$ remain irreducible.  This means that $\Omega$ is also
$\Spin_7^+$-invariant, whence the $\Spin_7^+$-orbit of any plane $\xi$
in the $\Omega$-grassmannian will also belong to the
$\Omega$-grassmannian. In fact, it is not difficult to show that the
$\Spin_7^+$-orbit {\em is\/} the $\Omega$-grassmannian, which in turn
coincides with the grassmannian of Cayley planes.  We will have more
to say about this below.

\subsubsection*{Eleven dimensions}

Now let $\Delta$ denote one of the two irreducible representations of
$\Cl_{1,10}$, and let $\psi\in\Delta$ be a spinor.  Squaring the
spinor we obtain on the right-hand side a 1-form $\Xi$, a 2-form
$\Psi$ and a 5-form $\Phi$:
\begin{equation}\label{eq:square}
\psi \otimes \bar\psi = \Xi + \Psi + \Phi~,
\end{equation}
where by $\bar\psi \equiv  -(e_0 \cdot \psi)^t$ we mean
the Majorana conjugate.  In this expression, the forms $\Xi$, $\Psi$
and $\Phi$ are respectively a $1$-, $2$- and $5$-form in
$\EE^{10,1}$.  Under the orthogonal decomposition $\EE^{10,1} =
\EE^{10} \oplus \RR e_0$, the 5-form $\Phi$ 
breaks up as
\begin{equation}\label{eq:bilinear}
\Phi = e_0^* \wedge \Lambda + \Theta~,
\end{equation}
where $\Lambda$ and $\Theta$ are a $4$- and a $5$-form on
$\EE^{10}$, respectively.  Now let $\xi$ be an oriented 5-plane in
$\EE^{10}$ and consider the bilinear $\bar\psi \xi \cdot
\psi$.  Using \eqref{eq:bilinear} and the definition of the
Majorana conjugate, one can rewrite this as
\begin{equation*}
\langle \psi, (e_0\wedge \xi )\cdot \psi\rangle =
\Theta(\xi) \tr \1 = 32\,\Theta(\xi)~,
\end{equation*}
where we have introduced the $\Spin_{10}$-invariant inner product
$\langle -,-\rangle$ defined by $\langle \chi,\psi\rangle =
\chi^t \psi$.  By the Cauchy--Schwarz inequality for this inner
product, we find that
\begin{equation}\label{eq:comass1}
\Theta(\xi) \leq \tfrac{1}{32} \|\psi\|
\|(e_0\wedge\xi)\cdot\psi\|~.
\end{equation}
Because $\xi$ is a unit simple 5-vector, $\|(e_0 \wedge\xi)\cdot\psi\|
= \|\psi\|$, whence
\begin{equation*}
\Theta(\xi) \leq \tfrac{1}{32} \|\psi\|^2.
\end{equation*}
In other words, the comass of $\Theta$ is given by
$\frac{1}{32}\|\psi\|^2$, and a 5-plane $\xi$ is calibrated by
$\Theta$ if and only if the $(5,1)$ plane $\pi = e_0\wedge\xi$
obeys $\pi \cdot \psi = \psi$, which is precisely the condition that
$\psi$ belongs to $\Delta(\pi)$.

The nature of the $\Theta$-grassmannian depends on the isotropy
group of the spinor $\psi$.  A nonzero Majorana spinor $\psi$ of
$\Spin_{10,1}$ can have two possible isotropy groups
\cite{Bryant-spinors}: either $\SU_5 \subset \Spin_{10}$, which 
acts trivially on a time-like direction which can be chosen to be
$e_0$, or a 30-dimensional non-semisimple Lie group $G \cong \Spin_7
\ltimes \RR^9$, acting trivially on a null direction.  In the former
case, the 5-form $\Theta$ is $\SU_5$-invariant and the
$\Theta$-grassmannian will contain the $\SU_5$-orbit of the plane
$\pi$.  This orbit turns out to be the full $\Theta$-grassmannian,
which is the grassmannian of special lagrangian planes in $\EE^{10}$.
In the latter case, $\Theta$ has the form $v^* \wedge \Omega$ where
$\Omega$ is a Cayley calibration in an eight-dimensional subspace $V
\subset \EE^{10}$ and $v \in V^\perp$ is a fixed vector perpendicular
to $V$.  In this case the $\Theta$-grassmannian agrees with the
$\Omega$-grassmannian, which is isomorphic to the grassmannian of
Cayley planes in $V\cong\EE^8$.

\subsection{Summary of results}

We can summarise the foregoing discussion as follows.  Given any
supersymmetric configuration of $\M5$-branes, the tangent planes
$\{\pi_i\}$ at any given singular point belong to a face of the
grassmannian: the intersection of the faces corresponding to the all
the spinors $\psi$ which belong to $\Delta(\pi_i)$ for all $i$.  We
will call such a face of the grassmannian, a {\em supersymmetric
face\/}.  The main problem in the study of the local singularity
structure of supersymmetric $\M5$-branes is the determination of the
supersymmetric faces of the grassmannian $G(5,1|10,1)$ of
$(5,1)$-planes in $\EE^{10,1}$, and for each such face to determine
the fraction $\nu$ of the supersymmetry which is preserved.

The first attempt at solving this problem was \cite{OhtaTownsend} who,
following up the work in \cite{Townsend-Mbranes}, classified the
supersymmetric static configurations of a pair of $\M5$-branes.  (Some
earlier incomplete results can be found in \cite{Jabbari}.) The
solution of the two fivebrane problem was completed in \cite{AFSS},
where we considered also fivebranes which are moving relative to each
other.  The multiple brane problem is still open, but some partial
results can be found in
\cite{GibbonsPapadopoulos,AFS-cali,AFS-groups,AFSS}.  As explained in
\cite{AFS-groups,AFSS}, but see also
\cite{BerkoozDouglasLeigh,GibbonsPapadopoulos,AFS-cali}, the
supersymmetric faces consist of planes which lie in the orbit of one
of the planes under the action of a subgroup of $\Spin_{10,1}$ which
leaves invariant some subspace of $\Delta$.  For each such subgroup
$G$ one can determine the fraction $\nu$ of the supersymmetry which is
preserved and the geometry defined by its orbit in the face of the
grassmannian.

We can distinguish two cases: faces in which all planes share a common
time-like direction and faces in which all planes share a common
light-like direction.   The former correspond to static brane
configurations, whereas the latter correspond to branes in motion.
Moreover, as shown in \cite{AFSS}, supersymmetric configurations of
brane are obtained by null-rotating (see, for example,
\cite{PenroseRindler}) already supersymmetric configurations
consisting of Cayley planes in eight dimensions.

\begin{table}[h!]
\centering
\setlength{\extrarowheight}{5pt}
\begin{tabular}{|*{6}{>{$}c<{$}|}}
\hline
\multicolumn{1}{|c|}{}&
\multicolumn{1}{c|}{Group}&
\multicolumn{1}{c|}{Isotropy}&
\multicolumn{1}{c|}{Geometry}&
\multicolumn{2}{c|}{Fraction $\nu$}\\
(p|n)&
G&
H&
G/H&
\text{static}&
\text{moving}\\[3pt]
\hline\hline
(5|10) & \SU_5 & \SO_5 & \slg_5 & \tfrac{1}{32}& -\\
(5|10) &\SU_2\times\SU_3 & \SO_2\times\SO_3& \slg_2 \times\slg_3 &
\tfrac{1}{16}& -\\ 
(4|8) &\Spin_7 &(\SU_2)^3/\ZZ_2 & \text{Cayley} & \tfrac{1}{32} &
\tfrac{1}{32}\\
(4|8) &\SU_4 & \SO_4 & \slg_4 & \tfrac{1}{16} & \tfrac{1}{32}\\
(4|8) &\SU_4 &\mathrm{S}(\U_2\times\U_2) & G_\CC(2|4) & \tfrac{1}{16}
& \tfrac{1}{16}\\
(4|8) &\Sp_2 &\U_2 & \clg_2 & \tfrac{3}{32} & \tfrac{1}{32},
\tfrac{1}{16}\\ 
(4|8) &\Sp_2 &\Sp_1\times\Sp_1 & G_\HH(1|2) & \tfrac{3}{32} &
\tfrac{3}{32}\\
(4|8) &\Sp_1\times\Sp_1 & \U_1\times\U_1 & \left(G_\CC(1|2)\right)^2 &
\tfrac{1}{8} & \tfrac{1}{16}\\
(4|8) &\Sp_1\times\Sp_1 & \Sp_1 & (3,1) & \tfrac{1}{8} &
\tfrac{1}{32}, \tfrac{3}{32}\\
(4|8) &\Sp_1 &\U_1 & (3,2) & \tfrac{5}{32} & \tfrac{1}{16},
\tfrac{3}{32}\\
(4|8) &\U_1 &\{1\} & (3,3) & \tfrac{3}{16} & \tfrac{3}{32}\\
(3|7) &G_2 &\SO_4 & \text{Associative} & \tfrac{1}{16} & \tfrac{1}{16}\\
(3|6) &\SU_3 &\SO_3 & \slg_3 & \tfrac{1}{8} & \tfrac{1}{16}\\
(2|6) &\SU_3 & \mathrm{S}(\U_2\times\U_1) & G_\CC(1|3) & \tfrac{1}{8}
& \tfrac{1}{8}\\
(2|4) &\SU_2 &\SO_2 &\slg_2 & \tfrac{1}{4} & \tfrac{1}{8}\\[3pt]
\hline
\end{tabular}
\vspace{8pt}
\caption{Some of the geometries associated with intersecting brane
configurations, together with the fraction of the supersymmetry which
is preserved both for static and for moving branes.}
\label{tab:summary}
\end{table}

We summarise the known results in Table~\ref{tab:summary}.  Each of
the geometries in the table is defined as the $G$-orbit of a $p$-plane
in $\EE^n$.  For each such geometry we list also the isotropy subgroup
$H\subset G$ of the reference $p$-plane, as well as the type of
(calibrated) geometry which one obtains.  We also tabulate the
fractions of supersymmetry both for static and (when $n\leq8$) moving
branes.  Some entries have more than one possible fraction $\nu$ for
moving branes.  These correspond to different but isomorphic subgroups
$G$.  The static fraction only depends on the conjugacy class of $G$
in $\Spin_{10}$, but the moving fraction is a more subtle invariant of
the configuration and depends intricately on how $G$ sits in
$\Spin_8$.

It may prove useful to explain one of the entries in detail.  Let us
consider for instance the fourth row in the table. These
configurations are obtained as follows.  For static configurations,
pick a $(5,1)$-plane $\pi = e_0 \wedge \xi$, where $\xi$ is a
$5$-plane in $e_0^\perp \cong \EE^{10}$.  The allowed configurations
consist of planes $\pi' = e_0 \wedge \xi'$, where $\xi'$ is in the
orbit of $\xi$ under a subgroup $G\cong \SU_4$.  $G$ leaves one
direction invariant, $v$ say, in $\xi$, so that the plane $\pi$ can be 
written as $\pi = e_0 \wedge v \wedge \zeta$, where $\zeta$ is a
$4$-plane in the eight-dimensional subspace of $e_0^\perp$ on which
$\SU_4$ acts irreducibly.  All other planes will be of the form $\pi'
= e_0 \wedge v \wedge \zeta'$ where $\zeta'$ is in the $G$-orbit of
$\zeta$.  The isotropy (in $G$) of $\zeta$ is a subgroup $H\cong\SO_4$
and with a little more effort one can recognise the subset of the
grassmannian as consisting of the special lagrangian $4$-planes.
Those configurations will generically preserve $\tfrac{1}{16}$ of the
supersymmetry.  For moving branes one simply starts with a
configuration of static branes, namely planes of the form $\pi' =
e_0\wedge v\wedge \zeta'$ where $\zeta'$ a special lagrangian
$4$-plane, and performs an arbitrary null rotation to each of the
planes.  Only null rotations along directions perpendicular to the
plane $\pi'$ change the configuration, whence the resulting
grassmannian is a homogeneous bundle over $G/H$ with fibre $\RR^5$.
The generic configuration now preserves $\tfrac{1}{32}$ of the
supersymmetry.

\section{Some geometries associated with intersecting branes}
\label{sec:geometries}

We now start a case-by-case description of the geometries in
Table~\ref{tab:summary}.  These geometries are not new, of course, but
some may not be well-known.  Complex geometries are of course
classical, and to some extent so are quaternionic geometries.  The
special lagrangian, associative and Cayley geometries were discussed
initially in Harvey \& Lawson's foundational essay on calibrated
geometry \cite{HarveyLawson}.  The complex lagrangian geometry (at
least in dimension eight) as well as the other geometries associated
to self-dual $4$-forms are discussed in \cite{DadokHarveyMorgan}.

\subsection{Complex geometry}\label{sec:complex}

The complex geometry of $k$-planes in $\CC^m\cong\RR^{2m}$ is defined
by the grassmannian $G_\CC(k|m) \subset G(2k|2m)$.  It is the
$\SU_m\subset\SO_{2m}$ orbit of a given real $2k$-plane.  Such planes
are calibrated by the properly normalised $k$th power of the K\"ahler
form
\begin{equation*}
\omega = \sum_{i=1}^m dx^i \wedge dy^i~,
\end{equation*}
where $z^i = x^i + \sqrt{-1} y^i$ are complex coordinates. It follows
from Wirtinger's inequality (see, for example, \cite{Federer}) that
the $2k$-form $\tfrac{1}{k!}\omega^k$ has unit comass and that its
grassmannian is precisely the grassmannian of complex $k$-planes.  The
K\"ahler form is left invariant by a $\U_m$ subgroup of $\SO_{2m}$,
whose intersection with the isotropy $\SO_{2k} \times \SO_{2m-2k}$ (in
$\SO_{2m}$) of a real plane is $\U_k \times \U_{m-k}$.  Note however
that the centre of $\U_m$ acts trivially, whence factoring it out, we
can write
\begin{equation*}
G_\CC(k|m) \cong
\frac{\SU_m}{\mathrm{S}\left(\U_k\times\U_{m-k}\right)}~.
\end{equation*}
In the study of $\M5$-branes we have the following grassmannians
appearing: $G_\CC(1|3)$, $G_\CC(2|4)\cong G(2|6)$, and
$G_\CC(1|2)\times G_\CC(1|2)\cong G(2|4)$.

\subsection{Quaternionic geometry}

Consider $\HH^\ell\cong\RR^{4\ell}$ and, on it, the quaternionic
$4$-form
\begin{equation*}
\Theta = \sum_{i=1}^\ell dx^i\wedge dy^i \wedge dz^i \wedge dw^i~,
\end{equation*}
where a quaternionic vector has components $q^i = x^i \ii + y^i \jj +
z^i \kk + w^i\1$.  Then results of Berger \cite{Berger} show that the
$4j$-form $\tfrac{6}{(2j+1)!}\Theta^j$ has unit comass, and the
corresponding grassmannian is nothing but the grassmannian
$G_{\HH}(j|\ell) \subset G(4j|4\ell)$ of quaternionic $j$-planes.
This grassmannian is acted on transitively by $\Sp_\ell \subset
\SO_{4\ell}$, and the intersection of $\Sp_\ell$ with the isotropy
$\SO_{4j}\times\SO_{4\ell -4j}$ of a real $4j$-plane, is given by
$\Sp_j\times \Sp_{\ell-j}$, whence
\begin{equation*}
G_\HH(j|\ell)\cong \frac{\Sp_\ell}{\Sp_j\times \Sp_{\ell-j}}~.
\end{equation*}
In the above table, it is $G_\HH(1|2) \cong G(1|5)\cong S^4$ which
appears.

\subsection{Special lagrangian geometry}

Special lagrangian geometry is another geometry associated to
$\SU_m\subset\SO_{2m}$.  This geometry is complementary to the
geometry of complex planes in $\CC^m\cong\RR^{2m}$.  Indeed, it is a
geometry of totally real planes.  Consider the forms
\begin{equation*}
\Lambda^{(\theta)} = \Re\, e^{i\theta} dz^1 \wedge dz^2 \wedge \cdots
\wedge dz^m~,
\end{equation*}
where $z^i$ are the complex coordinates for $\CC^m$ introduced in
Section~\ref{sec:complex} and $\theta \in S^1$.  It is shown in
\cite{HarveyLawson} that $\Lambda^{(\theta)}$ has unit comass, so that
it is a calibration.  Its grassmannian consists of the so-called {\em
special lagrangian\/} planes.  These planes are lagrangian with
respect to the K\"ahler form $\omega$ on $\CC^m$ defined in
Section~\ref{sec:complex}: that is, they are maximally isotropic
relative to $\omega$.  Notice however that the subset of $G(m|2m)$
consisting of {\em all\/} lagrangian planes (with respect to $\omega$)
is not the $\varphi$-grassmannian for any $\varphi$.  Nevertheless it
is fibred over the circle with fibres the special lagrangian planes
relative to $\Lambda^{(\theta)}$, for $\theta \in S^1$.  In other words,
every lagrangian plane is special lagrangian with respect to
$\Lambda^{(\theta)}$ for some $\theta$.  Notice that
$\U_m\subset\SO_{2m}$ does {\em not\/} preserve $\Lambda^{(\theta)}$
now, since the centre shifts $\theta$; but $\SU_m$ does.  Its
intersection with the isotropy $\SO_m\times\SO_m$ of an $m$-plane is
the diagonal $\SO_m$, whence the special lagrangian grassmannian
$\slg_m$ can be written as
\begin{equation*}
\slg_m \cong \frac{\SU_m}{\SO_m}~.
\end{equation*}
Notice that for $m{=}2$, special lagrangian geometries can be
identified with complex geometries relative to a complex structure
which is also left invariant by the same $\SU_2$ subgroup.  This is
because $\SU_2\cong\Sp_1$ actually leaves invariant a quaternionic
structure on $\HH\cong\CC^2$.  Other special lagrangian geometries
which appear in the table are $\slg_5$, $\slg_4 \cong G(3|6)$, and
$\slg_3$.

\subsection{Associative geometry}

Associative and Cayley geometries are intimately linked to the
octonions.  There are many constructions of the calibrations which
define these geometries, but they are all in one way or another
related to the octonions.  Let us therefore consider the $3$-form
$\varphi$ on $\RR^7$ defined as follows.  We identify
$\RR^7\cong\Im\OO$ with the imaginary octonions.  The octonions are a
normed algebra, whence in addition to a multiplication $\cdot$ they
also have an inner product $\langle,\rangle$.  The $3$-form $\varphi$
is defined by
\begin{equation*}
\varphi(a,b,c) = \langle a, b \cdot c\rangle~,
\end{equation*}
for all $a,b,c\in\Im\OO$.  We can choose a basis $o_i$,
$i=1,\ldots,7$, for the imaginary octonions, and canonically dual
basis $\theta_i$, relative to which $\varphi$ can be written as
\begin{equation*}
\varphi = \theta_{125} + \theta_{136} + \theta_{147} - \theta_{237} +
\theta_{246} - \theta_{345} + \theta_{567}~,
\end{equation*}
where we have used the shorthand $\theta_{ijk} \equiv \theta_i \wedge
\theta_j \wedge \theta_k$.  Harvey and Lawson proved in
\cite{HarveyLawson} that $\varphi$ is a calibration.  The
$\varphi$-grassmannian consists of so-called {\em associative\/}
planes, which are all constructed as follows.  Let $\ii,\jj,\kk$
generate any quaternion subalgebra of $\OO$.  Then the $3$-plane
$\ii\wedge \jj \wedge \kk$ is associative, and moreover all
associative planes are constructed in this way.

The group of automorphisms of the octonions is $G_2$, and its action
is such that it stabilises $\Im\OO$.  It also preserves the inner
product, whence it leaves $\varphi$ invariant.  In fact, $G_2$ can be
defined \cite{Bryant-holo} as the subgroup of $\GL_7\RR$ which leaves
$\varphi$ invariant.  The isotropy (in $G_2$) of an associative plane
is isomorphic to an $\SO_4$ subgroup, which acts on $\Im\OO$ as
follows.  We identify $\Im\OO$ with $\Im\HH \oplus \HH$ and $\SO_4$
with the $\ZZ_2$ quotient of $\Sp_1\times\Sp_1$, with $\Sp_1$ the unit
imaginary quaternions.  Thus if $g=(q_1,q_2)\in\Sp_1\times\Sp_1$, then
\begin{equation*}
g(a,b) = (q_1\cdot a\cdot \bar q_1, q_2\cdot b\cdot \bar q_1)~,
\end{equation*}
for $a\in\Im\HH$ and $b\in\HH$.  Notice that the $\ZZ_2$ subgroup
generated by $(-\1,-\1)\in\Sp_1\times\Sp_1$ acts trivially.  Clearly
$\ii \wedge \jj \wedge \kk$ is left invariant by $\SO_4$ and it is
shown in \cite{HarveyLawson} that $\SO_4$ is precisely the isotropy of
this $3$-plane.  In summary, the associative grassmannian is given by
\begin{equation*}
\text{Associative} \cong \frac{G_2}{\SO_4}~.
\end{equation*}

\subsection{Cayley geometry}

The Cayley grassmannian is the face exposed by a self-dual $4$-form
$\Omega$ in $\RR^8$, which we identify with $\OO$ as before.  Indeed,
we can build $\Omega$ in terms of the associative $3$-form $\varphi$
defined above, in the following way.  Consider the Hodge dual
$\Tilde\varphi$ (in $\Im\OO$) of $\varphi$:
\begin{equation*}
\Tilde\varphi \equiv \star_7 \varphi =  \theta_{1234} - \theta_{1267}
+ \theta_{1357} - \theta_{1456} + \theta_{2356} + \theta_{2457} +
  \theta_{3467}~,
\end{equation*}
in the obvious notation.  Thinking of $\Tilde\varphi$ as a $4$-form in
$\OO$, its Hodge dual is given by $\varphi\wedge \theta_8$, where
$\theta_8$ is the canonical dual form to $\1\in\OO$. We can now define
a self-dual 4-form $\Omega$ in $\OO$ as follows:
\begin{align*}
\Omega & = \Tilde\varphi + \varphi\wedge \theta_8\\
& = \theta_{1234} + \theta_{1258} - \theta_{1267} + \theta_{1357} +
\theta_{1368} - \theta_{1456} + \theta_{1478}\\
&\qquad {} + \theta_{2356} - \theta_{2378} + \theta_{2457} +
\theta_{2468} - \theta_{3458} + \theta_{3467} + \theta_{5678}~.
\end{align*}
As proven in \cite{HarveyLawson}, $\Omega$ has unit comass.  It is
known as the {\em Cayley calibration\/}, and its calibrated planes
make up the {\em Cayley grassmannian\/}.  Alternatively, $\Omega$ can
be defined in terms of the inner product on $\OO$ and the {\em triple
cross product\/}
\begin{equation*}
a \times b \times c = \half \left( a \cdot (\bar b \cdot c) - c\cdot
(\bar b \cdot a)\right)
\end{equation*}
as follows:
\begin{equation*}
\Omega(a,b,c,d) = \langle a \times b \times c, d\rangle~.
\end{equation*}
It follows that the typical calibrated plane is of the form $\1 \wedge
\ii \wedge \jj \wedge \kk$, where $\ii$, $\jj$, and $\kk=\ii\cdot\jj$
are the imaginary units in a quaternion subalgebra of $\OO$.

The Cayley form $\Omega$ is invariant under a $\Spin_7$ subgroup of
$\SO_8$, which acts transitively on the unit sphere in $\OO$ with
isotropy $G_2$.  As in the associative case, $\Spin_7$ can be defined
as the subgroup of $\GL_8\RR$ which leaves $\Omega$ invariant.  It
follows that $\Spin_7$ acts on the Cayley grassmannian.  This action
is transitive, with isotropy a subgroup $H \cong \left( \Sp_1 \times
\Sp_1 \times \Sp_1 \right)/\ZZ_2$ which acts as follows on $\OO$.  If
$g=(q_1,q_2,q_3)\in\Sp_1\times\Sp_1\times\Sp_1$ is a triple of unit
imaginary quaternions, then under $\OO=\HH\oplus\HH$ we have
\begin{equation*}
g (a,b) = (q_3 \cdot a \cdot \bar q_1, q_2 \cdot b \cdot \bar q_1)~,
\end{equation*}
for $a,b\in\HH$.  Notice that $(-1,-1,-1)$ acts trivially, whence the
action factors through $H$.  Clearly $H$ leaves $\1 \wedge \ii \wedge
\jj \wedge \kk$ invariant, and it is shown in \cite{HarveyLawson} that
$H$ is precisely the isotropy of such a plane.  In summary, the Cayley
grassmannian can be written as
\begin{equation*}
\text{Cayley} \cong \frac{\Spin_7}{H}~.
\end{equation*}
Notice that the Cayley grassmannian is isomorphic to $G(3|7)$.  This
is no accident, since given any oriented 3-plane in $\RR^7$, there is
a unique Cayley plane in $\RR^8$ which contains it.

\subsection{Complex lagrangian geometry}

The complex lagrangian geometry is a geometry of $2\ell$-planes in
$\RR^{4\ell}$.  Identifying $\RR^{4\ell}$ with $\HH^\ell$, determines
a quaternionic structure $I$, $J$, and $K=I\,J$.  The complex
lagrangian planes are those planes which are complex relative to $I$,
say, and lagrangian relative to $J$.  Let $\omega_I$ denote the
K\"ahler form relative to $I$, and $\Lambda_J^{(0)}$ denote the
special lagrangian form relative to $J$ with angle $\theta = 0$.  Then
consider the sum
\begin{equation*}
\Xi = \half \Lambda_J^{(0)} + \half \tfrac{1}{\ell!} \omega_I^\ell~.
\end{equation*}
One can show that $\Xi$ is a calibration, whose grassmannian consists
of those real $2\ell$-planes which are complex relative to $I$ and
(special) lagrangian relative to $J$; that is, the complex lagrangian
$2\ell$-planes.  The quaternionic structure $\{I,J,K\}$ determines an
$\Sp_\ell$ subgroup of $\SO_{4\ell}$, which leaves $\Xi$ invariant.
Its intersection with the isotropy of a reference complex lagrangian
plane is a $\U_\ell$ subgroup, whence
\begin{equation*}
\clg_\ell \cong \frac{\Sp_\ell}{\U_\ell}~.
\end{equation*}
Notice that $\clg_1\cong\slg_2\cong G_\CC(1|2)$.  Apart from this
degenerate case, it is $\clg_2\cong G(2|5)$ which appears in the
table.  In this case, it is not hard to show that $\Xi$ is actually
self-dual, as was the case for the Cayley, quaternionic and complex
geometries of real $4$-planes in $\RR^8$ discussed above, and for the
remaining three calibrations to be discussed below.

\subsection{Other geometries associated to self-dual $4$-forms}

It remains to discuss the three geometries labelled $(3,1)$, $(3,2)$
and $(3,3)$ in the table.  The notation has been borrowed from
\cite{DadokHarveyMorgan} who classified the (anti-)self-dual
calibrations in $\RR^8$, of which these are examples.

 Each one
in turn is associated to a certain self-dual calibration on $\RR^8$.
Let us choose an oriented basis $e_i$ for $\RR^8$ and let $\theta_i$
denote the canonical dual basis.  We will use the notation where
$e_{ijk\ell} = e_i\wedge e_j\wedge e_k\wedge e_\ell$ and similarly for
$\theta_{ijk\ell}$.  In addition let
\begin{equation*}
\theta^{ijk\ell} = \theta_{ijk\ell} + \star\,\theta_{ijk\ell}
\end{equation*}
be the manifestly self-dual extension of $\theta_{ijk\ell}$.  Consider
the following three self-dual forms
\begin{align}\label{eq:3iforms}
\Psi_{(3,1)} &= \theta^{1234} + \half \theta^{1256} + \half
\theta^{1467} - \half \theta^{1368}\notag\\
\Psi_{(3,2)} &= \theta^{1234} + \tfrac35 \theta^{1256} - \tfrac15
\theta^{1278} + \tfrac15 \theta^{1357} + \tfrac15 \theta^{1467} -
\tfrac15 \theta^{1368} + \tfrac15 \theta^{1458}\notag\\
\Psi_{(3,3)} &= \theta^{1234} + \tfrac13 \theta^{1256} - \tfrac13
\theta^{1368} + \tfrac13 \theta^{1458}~.
\end{align}
As shown in \cite{DadokHarveyMorgan} these forms have unit comass.  It
is clear from their explicit expressions that the $4$-plane $e_{1234}$
is calibrated by each of them.  These forms are left invariant by the
following subgroups of $\SO_8$:
\begin{align}\label{eq:3isotropy}
K_{(3,1)} &= \Sp_1 \cdot \left(\Sp_1 \times \Sp_1\right) \cong
\left(\Sp_1\times\Sp_1\times\Sp_1\right)/\ZZ_2 \notag\\
K_{(3,2)} &= \Sp_1 \cdot \left(\Sp_1 \times \U_1\right) \cong
\left(\Sp_1\times\Sp_1\times\U_1\right)/\ZZ_2\notag\\
K_{(3,3)} &= \Sp_1 \cdot \left(\U_1 \times \Sp_1\right) \cong
\left(\Sp_1\times\U_1\times\Sp_1\right)/\ZZ_2~,
\end{align}
which are all subgroups of the $\Sp_1\cdot\Sp_2 \cong
\left(\Sp_1\times\Sp_2\right)/\ZZ_2$ subgroup which leaves invariant
the quaternionic form
\begin{equation*}
\Theta = \theta^{1234} + \tfrac13 \theta^{1256} + \tfrac13
\theta^{1278} + \tfrac13 \theta^{1357} - \tfrac13 \theta^{1368} +
\tfrac13 \theta^{1458} + \tfrac13 \theta^{1467}~, 
\end{equation*}
which also calibrates $e_{1234}$.  This shows that these geometries
are subgeometries of the quaternionic geometry $G(\Theta) \cong
G_\HH(1|2)$.  As shown in \cite{DadokHarveyMorgan}, the grassmannians
$G\left(\Psi_{(3,i)}\right)$ coincide with the $K_{(3,i)}$ orbits of
$e_{1234}$.  Computing the intersection of the
$\SO_4\times\SO_4\subset\SO_8$ isotropy subgroup of $e_{1234}$ with
the $K_{(3,i)}$ and factoring out common subgroups, we obtain the
following description for the grassmannians
\begin{align*}
G\left(\Psi_{(3,1)}\right) &\cong \frac{\Sp_1\times\Sp_1}{\Sp_1}
\cong G(1|4) \cong S^3\\
G\left(\Psi_{(3,2)}\right) &\cong \frac{\Sp_1}{\U_1}
\cong G(1|3) \cong S^2\\
G\left(\Psi_{(3,1)}\right) &\cong \U_1 \cong G(1|2) \cong S^1~.
\end{align*}

\section{The eight-dimensional geometries in detail}
\label{sec:tour}

In this section we will go in more detail through the
eight-dimensional geometries in Table \ref{tab:summary}---that is, the
subgeometries of $G(4|8)$.  There are nine such geometries: Cayley,
complex (two kinds), quaternionic, special lagrangian, complex
lagrangian, as well as the $(3,i)$ subgeometries of the quaternionic
geometry.  All these geometries share the property that they are
calibrated by self-dual $4$-forms in $\RR^8$.  The strategy in this
section is the following.  We fix a given $4$-plane in $\EE^8$ and we
will describe the orbits of this plane under different subgroups of
$\SO_8$.  In many cases, these subgroups will be determined uniquely
by specifying a certain structure (complex, quaternionic,...) in
$\EE^8$ which it leaves invariant.

\subsection{Notation and basic strategy}

We will let $\{e_i\}$ for $i=1,2,\ldots,8$ be an oriented orthonormal
basis for $\EE^8$, and introduce the shorthand notation $e_{ij...k} =
e_i\wedge e_j \wedge \cdots \wedge e_k$.  This choice of basis allows
us to identify $\EE^8$ with its dual, and forms with polyvectors.  Our
reference oriented $4$-plane will be $e_{1234}$.  Its $\SO_8$-isotropy
$K$ is isomorphic to $\SO_4 \times \SO_4$, the first factor acting on
the span of $e_{1234}$ and the second on the span of $e_{5678}$.  The
grassmannian of oriented $4$-planes in $\EE^8$ is then the
$\SO_8$-orbit of $e_{1234}$,
\begin{equation*}
G(4|8) = \SO_8 \cdot e_{1234} \cong \frac{\SO_8}{\SO_4\times\SO_4}~.
\end{equation*}
For every subgroup $G \subset \SO_8$, the $G$-orbit of $e_{1234}$ is a
subset of the grassmannian which is itself isomorphic to a coset
space,
\begin{equation*}
G(4|8) \supset G \cdot e_{1234} \cong \frac{G}{G \cap K}~.
\end{equation*}
In what follows we will specify the group $G$ in terms of invariant
structures on $\EE^8$.

It is well-known that a complex structure determines an $\SU_4$
subgroup of $\SO_8$ which shares its maximal torus with a $\Spin_7$
subgroup.  Also a quaternionic structure determines an $\Sp_2$
subgroup of $\SO_8$.  This $\Sp_2$ subgroup is nothing but the
intersection of the $\SU_4$ subgroups corresponding to each of the
three complex structures in the quaternionic structure.  A
quaternionic structure allows us to think of $\EE^8$ as $\HH^2$.  A
given split $\HH^2 = \HH \oplus \HH$ is preserved by an $\Sp_1\times
\Sp_1$ subgroup of the $\Sp_2$, and this $\Sp_1\times\Sp_1$ subgroup
in turn determines a diagonal $\Sp_1$ subgroup, whose maximal
torus defines a $\U_1$ subgroup.  Starting with different complex
structures and some extra structure along the way, we will therefore
be able to construct all the geometries of interest.

\subsection{A guided tour}

We start, following \cite{DadokHarveyMorgan}, by introducing a
convenient notation for complex structures in $\EE^8$.  By a complex
structure
\begin{equation}\label{eq:i}
I = \begin{pmatrix}
    1 & 3 & 5 & 7\\
    2 & 4 & 6 & 8
    \end{pmatrix}~,
\end{equation}
we mean that $I\,e_1 = e_2$, $I\,e_2 = -e_1$, $I\,e_3 = e_4$, etc.
Each complex structure determines a ``K\"ahler'' $2$-form, which in
this case is given by
\begin{equation*}
\omega_I = e_{12} + e_{34} + e_{56}+ e_{78}~,
\end{equation*}
which in turn defines a self-dual $4$-form, called the K\"ahler
calibration:
\begin{equation}\label{eq:kaehlerform}
\half \omega_I^2 = e^{1234} + e^{1256} + e^{1278}~,
\end{equation}
where as above we have introduced the explicit self-dual $4$-forms
\begin{equation*}
e^{ijkl} = e_{ijkl} + \star e_{ijkl}~.
\end{equation*}

A complex structure $I$ also defines a special lagrangian calibration
$\Lambda_I$ in the following way.  We start by defining the following
complex vectors:
\begin{equation*}
\zeta_i = e_{2i-1} + \sqrt{-1}\,e_{2i}~,
\end{equation*}
for $i=1,2,3,4$.  They have the virtue that they are eigenvectors of
$I$ and therefore ``diagonalise'' the K\"ahler form:
\begin{equation*}
\omega_I = \half \Im \sum_{i=1}^4 \zeta_i \wedge \bar\zeta_i~.
\end{equation*}
The special lagrangian calibration $\Lambda_I$ is then defined as the
following real $4$-form:
\begin{equation*}
\Lambda_I = \Re \left( \zeta_1 \wedge \zeta_2 \wedge \zeta_3 \wedge
\zeta_4 \right)~,
\end{equation*}
expanding to
\begin{equation*}
\Lambda_I = e^{1357} - e^{1368} - e^{1467} - e^{1458}~,
\end{equation*}
which is manifestly self-dual.  Notice that $\Lambda_I$ does not
calibrate $e_{1234}$.  This is to be expected because a plane cannot
be both complex and lagrangian (hence totally real) relative to the
same complex structure.

Therefore we choose a second complex structure $J$ defined by
\begin{equation}\label{eq:j}
J = \begin{pmatrix}
    1 & 2 & 3 & 4\\
    8 & 7 & 6 & 5
    \end{pmatrix}~.
\end{equation}
Its K\"ahler form is given by $\omega_J = e_{18} + e_{27} + e_{36} +
e_{45}$, which squares to
\begin{equation}\label{eq:jsqrd}
\half \omega_J^2 = e^{1278} + e^{1368} + e^{1458}~.
\end{equation}
The special lagrangian form is given by
\begin{equation}\label{eq:slagform}
\Lambda_J = e^{1234} + e^{1256} - e^{1357} + e^{1467}~,
\end{equation}
which clearly calibrates $e_{1234}$.  Whereas the special lagrangian
calibration $\Lambda_J$ is $\SU_4$-invariant, the K\"ahler calibration
$\half\omega_J^2$ is actually $\U_4$-invariant.  Nevertheless the
centre of $\U_4$, being generated by the complex structure $J$ itself,
stabilises the plane, whence, just as for the special lagrangian
grassmannian, the complex grassmannian is an $\SU_4$ orbit.

Now consider the combination
\begin{equation*}
\Omega_J = \Lambda_J - \half \omega_J^2~.
\end{equation*}
As shown in \cite{HarveyLawson}, this is a Cayley form and is left
invariant by the $\Spin_7$ subgroup of $\SO_8$ which contains (and
shares the same maximal torus with) the $\SU_4$ leaving $J$ invariant.
In our case, $\Omega_J$ expands to
\begin{equation}\label{eq:cayleyform}
\Omega_J = e^{1234} + e^{1256} - e^{1278} - e^{1357} - e^{1368} -
e^{1458} + e^{1467}~,
\end{equation}
from which we see that it calibrates $e_{1234}$.

The two complex structures $I$ and $J$ defined above anticommute: $K
\equiv IJ = -JI$, where
\begin{equation}\label{eq:k}
K = \begin{pmatrix}
    \phantom{-}1 & 2 & \phantom{-}3 & 4\\
    -7 & 8 & -5 & 6
    \end{pmatrix}~,
\end{equation}
correcting a typo in \cite{DadokHarveyMorgan}.  Therefore $\{I,J,K\}$
define a quaternionic structure on $\EE^8$.  The intersection of the
$\SU_4$ subgroups corresponding to the three complex structures is an
$\Sp_2$ subgroup of $\SO_8$.  Given an $\Sp_2$ subgroup it gives rise
to a family of 24 quaternionic structures: all possible reorderings
and consistent sign changes in $\{I,J,K\}$.  The K\"ahler and special
lagrangian calibrations for each of the complex structures in the
quaternionic structure satisfy a number of useful identities:
\begin{align}\label{eq:ides}
\Lambda_I &= \half \omega_K^2  - \half \omega_J^2\notag\\
\Lambda_J &= \half \omega_I^2  - \half \omega_K^2\notag\\
\Lambda_K &= \Lambda_I + \Lambda_J~.
\end{align}

A useful way to construct new calibrations out of old ones is to take
{\em convex\/} linear combinations.  By this we mean a linear
combination $\sum_i a_i C_i$, where each $C_i$ is a calibration and
$a_i \geq 0$ with $\sum_i a_i = 1$.  Such a linear combination is
automatically a calibration and moreover its grassmannian is the
intersection of the $C_i$-grassmannians.  Because $e_{1234}$ belongs
to both the complex grassmannian corresponding to $I$ and to the
special lagrangian grassmannian corresponding to $J$, we can take the
following convex linear combination
\begin{equation*}
\Xi = \tfrac14 \omega_I^2 + \half \Lambda_J~,
\end{equation*}
which expands to
\begin{equation}\label{eq:clagform}
\Xi = e^{1234} + e^{1256} + \half e^{1278} - \half e^{1357} + \half
e^{1467}~.
\end{equation}
Its grassmannian consists of those planes which are complex with
respect to $I$ and special lagrangian with respect to $J$.  The
resulting geometry is called complex lagrangian.  The same geometry
arises as the calibrated geometry of the convex linear combination
\begin{equation*}
\Xi' = \half \Lambda_J + \half \Lambda_K~,
\end{equation*}
which expands to
\begin{equation}\label{eq:clagformtoo}
\Xi' = e^{1234} + e^{1256} - \half e^{1357} - \half e^{1368} - \half
e^{1458} + \half e^{1467}~.
\end{equation}
The $\Xi'$-grassmannian consists of planes which are special
lagrangian with respect to both $J$ and $K$.  It is not hard to show
that the $\Xi'$- and $\Xi$-geometries agree.

Indeed, it is enough to show that if $\xi$ is special lagrangian with
respect to $J$, then $\xi$ is special lagrangian with respect to $K$
if and only if it is complex with respect to $I$.  Using the fact that
for any complex structure, the K\"ahler calibration $\half\omega^2$ is
identically zero on the special lagrangian grassmannian $G(\Lambda)$,
and the first identity in \eqref{eq:ides}, it follows that $\Lambda_I$
and $\half\omega_K^2$ agree on $G(\Lambda_J)$.  Therefore if a plane
$\xi$ in $G(\Lambda_J)$ is also in $G(\Lambda_K)$ then
$\half\omega_K^2(\xi)=0$, whence $\Lambda_I(\xi)=0$ so that $\xi\in
G(\half\omega_I^2)$. Similarly if $\xi$ is in $G(\half\omega_I^2)$,
then $\Lambda_I(\xi)=0$ whence $\half\omega_K^2(\xi)=0$ and $\xi\in
G(\Lambda_K)$.

A useful convex linear combination of calibrations is the quaternionic
calibration.  Given a quaternionic structure $\{I,J,K\}$, we can
define a quaternionic $4$-form
\begin{equation*}
\Theta_{\{I,J,K\}} \equiv \tfrac16 \left( \omega_I^2 + \omega_J^2 +
\omega_K^2 \right)~.
\end{equation*}
Being a convex linear combination of K\"ahler calibrations,
$\Theta_{\{I,J,K\}}$ is also a calibration whose grassmannian consists
of planes which are complex with respect to each of the complex
structures $I$, $J$ and $K$.  For this reason $e_{1234}$, being a
special lagrangian plane relative to $J$ cannot be quaternionic
relative to $\Theta$.  We remedy this by defining another complex
structure
\begin{equation*}
J' = \begin{pmatrix}
    1 & \phantom{-}2 & 5 & \phantom{-}6\\
    3 & -4 & 7 & -8
    \end{pmatrix}~,
\end{equation*}
which also anticommutes with $I$.  Therefore $\{I'=I,J',K'=I'J'\}$
define a quaternionic structure, with quaternionic form $\Theta \equiv
\Theta_{\{I',J',K'\}}$ given by
\begin{equation}\label{eq:quatform}
\Theta = e^{1234} + \tfrac13 e^{1256} + \tfrac13 e^{1278} + \tfrac13
e^{1357} - \tfrac13 e^{1368} + \tfrac13 e^{1458} + \tfrac13 e^{1467}~,
\end{equation}
which now clearly calibrates $e_{1234}$.  As in the case of the
K\"ahler calibration, $\Theta$ is actually invariant under
$\Sp_1\cdot\Sp_2$; but because the $\Sp_1$ factor is generated by the
quaternionic structure itself, the quaternionic grassmannian is
actually the $\Sp_2$-orbit of $e_{1234}$.

In contrast to a quaternionic structure, which consists of two
anticommuting complex structures, let us consider two {\em
commuting\/} complex structures: $I$ defined in \eqref{eq:i} and $I''$
defined by
\begin{equation}\label{eq:ipp}
I'' = \begin{pmatrix}
     1 & \phantom{-}3 & \phantom{-}5 & 7\\
     2 & -4 & -6 & 8
    \end{pmatrix}~.
\end{equation}
Let us consider the self-dual form (again correcting a typo in
\cite{DadokHarveyMorgan})
\begin{equation}\label{eq:bicomp}
\Sigma \equiv \tfrac14 \omega_I^2 - \tfrac14 \omega_{I''}^2
= e^{1234} + e^{1256} = \left(e_{12} + e_{78}\right) \wedge
\left(e_{34} + e_{56}\right)~.
\end{equation}
In order to see that this form is a calibration, it is easiest to
rewrite it as a convex linear combination of special lagrangian
forms
\begin{equation*}
\Sigma = \half \Lambda_J + \half \Lambda_{J''}~,
\end{equation*}
where $J$ is the complex structure in \eqref{eq:j} and $J''$ is given
by
\begin{equation*}
J'' = \begin{pmatrix}
       \phantom{-}1 & \phantom{-}2 & 3 & 4\\
       -8 & -7 & 6 & 5
       \end{pmatrix}~,
\end{equation*}
which corrects yet another typo in \cite{DadokHarveyMorgan}.  The
complex structures $J$ and $J''$ are also commuting. Moreover, $I''$
and $J''$ anticommute, whence $\{I,J,K=IJ\}$ and $\{I'',J'',K'' =
I''J''\}$ are two commuting quaternionic structures.

From the product form of $\Sigma$ in \eqref{eq:bicomp}, we see that
the $\Sigma$-planes are products of a complex plane in the span of
$e_{1278}$ relative to $I_1 = \begin{pmatrix} 1 & 7\\ 2& 8
\end{pmatrix}$ and a complex plane relative to $I_2 = \begin{pmatrix}
3 & 5\\ 4 & 6 \end{pmatrix}$ in the span of $e_{3456}$.
Equivalently\footnote{This exemplifies the fact that $2$-planes in
$\EE^4$ which are complex relative to a K\"ahler calibration
$\omega_I$, are special lagrangian relative to $\Lambda_J$, where $I$,
$J$, and $K=IJ$ defines a quaternionic structure.  This is because of
the isomorphism $\SU_2 \cong \Sp_1$, so that the $\SU_2$ which leaves
$I$ invariant actually leaves invariant a quaternionic structure.},
$\Sigma$-planes are products of a special lagrangian plane relative to
$J_1 = \begin{pmatrix} 1 & 2\\ 8 & 7\end{pmatrix}$ in the span of
$e_{1278}$ and a special lagrangian plane relative to $J_2 =
\begin{pmatrix} 3 & 4\\ 6 & 5\end{pmatrix}$ in the span of
$e_{3456}$.  $\Sigma$ is invariant under an $\U_2\times\U_2$ subgroup
of $\SO_8$.  The centre stabilises $e_{1234}$---in fact, it stabilises
the spans of $e_{12}$ and $e_{34}$ separately---whence the
$\Sigma$-grassmannian is the $\Sp_1\times\Sp_1$ orbit of $e_{1234}$.

Finally we point out that the calibrations corresponding to the
$(3,i)$ geometries can be constructed out of complex and quaternionic
structures.  In fact, we have the following expressions
\begin{align*}
\Psi_{(3,1)} &= \tfrac34 \Theta + \tfrac14 \Omega_J\\
\Psi_{(3,2)} &= \tfrac35 \Theta - \tfrac25 \omega^2_{I''}\\
\Psi_{(3,3)} &= \half \Theta - \half \Tilde\Theta~,
\end{align*}
where $J$ and $I''$ are the complex structures defined by \eqref{eq:j}
and \eqref{eq:ipp} respectively, and where $\Tilde\Theta \equiv
\Theta_{\{\Tilde I, \Tilde J, \Tilde K\}}$, given by
\begin{equation*}
\Hat\Theta = - e^{1234} - \tfrac13 e^{1256} + \tfrac13 e^{1278} +
\tfrac13 e^{1357} + \tfrac13 e^{1368} - \tfrac13 e^{1458} + \tfrac13
e^{1467}~,
\end{equation*}
is the quaternionic calibration corresponding to the quaternionic
structure generated by $\Tilde I = I''$ in \eqref{eq:ipp} and
\begin{equation*}
\Tilde J = \begin{pmatrix}
	   1 & 2 & 5 & 6\\
           3 & 4 & 7 & 8
	   \end{pmatrix}~.
\end{equation*}

\section{Generalised self-duality}
\label{sec:selfdual}

Every (constant coefficient) $4$-form $\varphi$ in $\EE^8$ defines an
endomorphism of the space of $2$-forms:
\begin{align}\label{eq:2formmap}
\Hat\varphi : {\textstyle\bigwedge^2 \EE^8} &\to
 {\textstyle\bigwedge^2 \EE^8}\notag\\
\omega &\mapsto \star \left(\star\varphi \wedge \omega\right)~.
\end{align}
Explicitly, if $\varphi = \sum_{i<j<k<l} \varphi_{ijkl}e_{ijkl}$ and
$\omega = \sum_{i<j} \omega_{ij} e_{ij}$ then
\begin{equation*}
(\Hat\varphi\, \omega)_{ij} = \sum_{k<l} \varphi_{ijkl}\,\omega_{kl}~.
\end{equation*}
This expression clearly shows that $\Hat\varphi$ is traceless, and
symmetric under the natural inner product
\begin{equation*}
\langle \alpha, \beta\rangle \equiv \star \left( \alpha
\wedge\star\beta\right) = \sum_{i<j} \alpha_{ij}\,\beta_{ij}
\end{equation*}
on the space of $2$-forms.  This means that $\Hat\varphi$ will be
diagonalisable.  If $G$ is the $\SO_8$-isotropy subgroup of $\varphi$,
then the eigenspaces of $\Hat\varphi$ are $G$-submodules of
$\bigwedge^2 \EE^8$, the $\repre{28}$ or adjoint representation
$\so_8$ of $\SO_8$.  A canonical $G$-submodule is the adjoint
representation $\fg \subset \so_8$, but of course there are other
$G$-submodules as well.  One can use $\Hat\varphi$ to define a
generalised {\em self-duality\/} for $2$-forms in eight dimensions by
demanding that a $2$-form belong to a definite $G$-submodule of
$\so_8$.  This generalises self-duality in four dimensions, where we
can take $\varphi = \star 1$, and $\Hat\varphi = \star$ itself.  The
eigenspaces of $\Hat\varphi$ in this case are the subspaces of
self-dual and anti-self-dual $2$-forms: corresponding to the adjoint
representations of the two $\Sp_1$ factors in $\SO_4 \cong
\Sp_1\cdot\Sp_1$.

Generalised self-duality plays a crucial role in the attempts to
generalise the notion of Yang--Mills instantons to higher dimensions
\cite{CDFN,Ward}.  Suppose that the Yang--Mills curvature $F(A)$
satisfies a generalised self-duality condition
\begin{equation}\label{eq:gsdc}
\Hat\varphi\,F(A) = c\,F(A)~,
\end{equation}
for some {\em nonzero\/} constant $c$.  Then one easily computes
\begin{align*}
d_A\star F(A) &=  c^{-1} d_A\,\left( \star\varphi\wedge
F(A)\right)\\
&= c^{-1} \left( d\star\varphi \wedge F(A) + \star\varphi \wedge
d_A\,F(A)\right)~,
\end{align*}
whence using the Bianchi identity $d_A F(A)=0$ and {\em provided that
$\star\varphi$ is closed\/}, the Yang--Mills equations of motion are
satisfied automatically.  In the geometries under consideration
$\varphi$ is self-dual and it is constant, so that it is co-closed.

In what follows we will discuss the possible notions of self-duality
which are available for each of the above geometries in eight
dimensions, by analysing the eigenspace decompositions of the
endomorphisms $\Hat\varphi$ corresponding to the different
calibrations $\varphi$ described above.  We should remark however that
despite the fact that a one-to-one correspondence between geometries
and generalised self-duality conditions is not expected---after all
self-duality depends crucially on the calibration, whereas as we saw
above for the case of the complex lagrangian geometry, different
calibrations can give rise to the same geometry---nevertheless we will
see that in some cases the geometry does determine the possible
generalised self-dualities.

\subsection{Cayley geometry}

The Cayley calibration \eqref{eq:cayleyform} is invariant under a
$\Spin_7$-subgroup of $\SO_8$, under which the $\repre{28}$ breaks up as
\begin{equation*}
\repre{28} \to \repre{7} \oplus \repre{21}~,
\end{equation*}
where the $\repre{21}$ corresponds to the adjoint representation
$\spin_7 \subset \so_8$.  It is well known that the endomorphism
$\Hat\Omega$ obeys the following characteristic polynomial:
\begin{equation*}
\left( \Hat\Omega -\1\right) \left( \Hat\Omega +3\1\right) = 0~,
\end{equation*}
whence we see that the eigenvalues are $1$ and $-3$, and (using
tracelessness of $\Hat\Omega$) with multiplicities $21$ and $7$,
respectively.  Therefore there are two possible notions of
self-duality, and hence two possible extensions of the notion of
instanton to eight dimensions.  As shown in \cite{AFOS}, supersymmetry
seems to prefer the definition of instanton which says
that $F(A)$ belongs to $\spin_7\subset\so_8$: $\Hat\Omega F(A) =
F(A)$.  Gauge fields satisfying this relation are known as {\em
octonionic instantons\/}, for reasons explained in \cite{Moduli}.

\subsection{Complex geometries}

Let $I$ denote the complex structure defined in equation
\eqref{eq:i}, and let $\Upsilon \equiv \half\omega^2_I$, which is
given by \eqref{eq:kaehlerform}.  Let $\Hat\Upsilon$ denote the
endomorphism of $2$-forms defined from $\Upsilon$ according to
\eqref{eq:2formmap}.  Its characteristic polynomial is given by
\begin{equation*}
\left(\Hat\Upsilon - 3\1\right) \left(\Hat\Upsilon - \1\right)
\left(\Hat\Upsilon + \1\right) = 0~,
\end{equation*}
whence $\Upsilon$ has three eigenvalues $3$, $1$ and $-1$.  The
multiplicities are $1$, $12$ and $15$ respectively.  $\Upsilon$ is
$\U_4$ invariant, and under $\U_4 \cong \left(\SU_4\times
\U_1\right)/\ZZ_4$ the $\repre{28}$ breaks up as
\begin{equation*}
\repre{28} \to \repre{6}_2 \oplus \repre{6}_{-2} \oplus \repre{15}_0
\oplus \repre{1}_0~,
\end{equation*}
where the last two factors correspond to the adjoint representation
$\u_4 = \su_4 \oplus \u_1 \subset \so_8$, and where the first two
factors together make up an irreducible {\em real\/} representation of
dimension $12$.  Therefore there is no accidental degeneracy in the
eigenspace decomposition of $\Hat\Upsilon$, in the sense that the
group theory does not refine any further the eigenvalues of
$\Hat\Upsilon$.  The natural self-duality condition in gauge theory
is the one which says that $F(A)$ belongs to $\su_4\subset\so_8$:
$\Hat\Upsilon F(A) = - F(A)$.  These equations are the well-known
K\"ahler--Yang--Mills equations, studied in \cite{D,UY}.

\subsection{Special lagrangian geometry}

Let $\Lambda \equiv \Lambda_J$ denote the special lagrangian form
defined by equation \eqref{eq:slagform}.  It is invariant under
$\SU_4\subset\SO_8$, under which the $\repre{28}$ breaks up as
\begin{equation*}
\repre{28} \to 2\,\repre{6} \oplus \repre{1} \oplus \repre{15}~,
\end{equation*}
where now each $\repre{6}$ is a real representation of
$\SU_4\cong\Spin_6$.  The map $\Hat\Lambda$ on $2$-forms obeys the
following characteristic polynomial:
\begin{equation*}
\left(\Hat\Lambda + 2\1\right) \Hat\Lambda \left(\Hat\Lambda -
2\1\right) = 0~,
\end{equation*}
whence it has eigenvalues $-2$, $0$ and $2$.  The multiplicities can
easily worked to be $6$, $16$ and $6$, which shows that the eigenvalue
$0$ is degenerate.

\subsection{Complex lagrangian geometry}

Let $\Xi$ denote the complex lagrangian calibration given by
\eqref{eq:clagform}.  The corresponding endomorphism $\Hat\Xi$
satisfies the characteristic polynomial
\begin{equation*}
\left(\Hat\Xi - \tfrac52 \1\right) \left(\Hat\Xi - \tfrac32 \1\right)
\left(\Hat\Xi - \half \1\right) \left(\Hat\Xi + \half \1\right)
\left(\Hat\Xi + \tfrac32 \1\right) = 0~,
\end{equation*}
so that it has five eigenvalues: $\tfrac52$, $\tfrac32$, $\half$,
$-\half$ and $-\tfrac32$.  The multiplicities are $1$, $5$, $6$, $11$
and $5$, which again agrees with $\Hat\Xi$ being traceless.  The
eigenvalues $\pm\half$ are now degenerate, a fact for which there
seems to be no group-theoretical explanation, since $\Xi$ is precisely
$\Sp_2$-invariant, and under $\Sp_2$ the $\repre{28}$ breaks up as
\begin{equation*}
\repre{28} \to 3\,\repre{1} \oplus 3\,\repre{5} \oplus \repre{10}~.
\end{equation*}
The three singlets correspond to $\omega_I$, $\omega_J$ and
$\omega_K$, and the $\repre{10}$ corresponds to the adjoint
representation $\sp_2 \subset \so_8$.

Similarly, let $\Xi'$ be the other complex lagrangian calibration
defined by \eqref{eq:clagformtoo} and let $\Hat\Xi'$ be the
corresponding endomorphism.  It satisfies the characteristic
polynomial
\begin{equation*}
\left(\Hat\Xi' + 2 \1\right) \left(\Hat\Xi' + \1\right) \Hat\Xi'
\left(\Hat\Xi' - \1\right) \left(\Hat\Xi' - 2 \1\right) = 0~.
\end{equation*}
The five eigenvalues $-2$, $-1$, $0$, $1$ and $2$ have multiplicities
$5$, $2$, $10$, $10$ and $1$ respectively.  $\Hat\Xi'$ is actually
invariant under an $\U_1\cdot\Sp_2 =
\left(\U_1\times\Sp_2\right)/\ZZ_2$ subgroup of $\SO_8$.  Under this
subgroup the $\repre{28}$ breaks up as
\begin{equation}\label{eq:bilagdec}
\repre{28} \to \repre{1}_0 \oplus \repre{10}_0 \oplus
\repre{5}_0\oplus \repre{1}_{2} \oplus \repre{1}_{-2} \oplus
\repre{5}_{2} \oplus \repre{5}_{-2}~.
\end{equation}
The first two factors correspond to the adjoint representation
$\u_1\oplus\sp_2\subset \so_8$.  In terms of real representations,
$\repre{1}_{2} \oplus \repre{1}_{-2}$ is an irreducible
$2$-dimensional representation and $\repre{5}_{2} \oplus
\repre{5}_{-2}$ is an irreducible $10$-dimensional
representation. Therefore there is no degeneracy in the spectrum of
$\Hat\Xi'$.

\subsection{Quaternionic geometry}

The linear map $\Hat\Theta$ associated to the quaternionic $4$-form
$\Theta$ in \eqref{eq:quatform} obeys the characteristic polynomial
\begin{equation*}
\left(\Hat\Theta + \1\right)\left(\Hat\Theta - \tfrac13 \1\right)
\left(\Hat\Theta - \tfrac53 \1\right) = 0~,
\end{equation*}
so that it has three eigenvalues: $-1$, $\tfrac13$ and $\tfrac53$.
The multiplicities are $10$, $15$ and $3$ respectively, reiterating
the fact that $\Hat\Theta$ is traceless.  The degeneracy of the
eigenvalues is easily explained if we remark that $\Theta$ is actually
invariant under the maximal subgroup $\Sp_1\cdot \Sp_2 =
\left(\Sp_1\times\Sp_2\right)/\ZZ_2$ of $\SO_8$, under which the
$\repre{28}$ decomposes into three factors as
\begin{equation*}
\repre{28} \to (\repre{3},\repre{1}) \oplus (\repre{1},\repre{10})
\oplus (\repre{3},\repre{5})~,
\end{equation*}
the first two factors corresponding to the adjoint representation
$\sp_1 \oplus \sp_2 \subset \so_8$.  The corresponding self-duality
equations for Yang--Mills fields were originally studied, in the
context of quaternionic K\"ahler manifolds, in
\cite{MCS,GalickiPoon}.

\subsection{Sub-quaternionic geometries}

Finally let us consider the self-dual forms defined by
\eqref{eq:3iforms}.  Let $\Hat\Psi_i$ denote the endomorphism of
$2$-forms defined by the calibration $\Psi_{(3,i)}$.  These maps obey
the following characteristic polynomials
\begin{align*}
\left(\Hat\Psi_1 - \tfrac32 \1\right)
\left(\Hat\Psi_1 - \half \1\right)
\left(\Hat\Psi_1 + \half \1\right)
\left(\Hat\Psi_1 + \tfrac32 \1\right) &= 0\\
\left(\Hat\Psi_2 - \tfrac75 \1\right)
\left(\Hat\Psi_2 - \tfrac35 \1\right)
\left(\Hat\Psi_2 + \tfrac15 \1\right)
\left(\Hat\Psi_2 + \1\right)
\left(\Hat\Psi_2 + \tfrac95 \1\right) &= 0\\
\left(\Hat\Psi_3 - \tfrac43 \1\right)
\left(\Hat\Psi_3 - \tfrac23 \1\right)
\Hat\Psi_3
\left(\Hat\Psi_3 + \tfrac23 \1\right)
\left(\Hat\Psi_3 + \tfrac43 \1\right) &=0~.
\end{align*}
The multiplicities of the eigenvalues are given as follows: for
$\Hat\Psi_1$ the eigenvalues are $\tfrac32$, $\half$, $-\half$, and
$-\tfrac32$ with multiplicities $3$, $12$, $9$ and $4$ respectively;
for $\Hat\Psi_2$ the eigenvalues are $\tfrac75$, $\tfrac35$,
$-\tfrac15$, $-1$ and $-\tfrac95$ and $-\tfrac32$ with multiplicities
$3$, $9$, $9$, $6$ and $1$ respectively; and for $\Hat\Psi_3$ the
eigenvalues are $\tfrac43$, $\tfrac23$, $0$, $-\tfrac23$ and
$-\tfrac43$ with multiplicities $3$, $6$, $10$, $6$ and $3$
respectively.  The forms $\Psi_{(3,i)}$ are invariant with respect to
the subgroups $K_{(3,i)}$ of $\SO_8$ given in equation
\eqref{eq:3isotropy}.  As mentioned above, these groups are subgroups
of the $\Sp_1\cdot\Sp_2$ isotropy of the quaternionic form $\Theta$ in
equation \eqref{eq:quatform}.  In fact, the first $\Sp_1$ factor in
$K_{(3,i)}$ is precisely the same as the one in $\Sp_1\cdot\Sp_2$.
All eigenspace decompositions are degenerate for these three groups.
As an example, let us work out the $(3,1)$ geometry.  Under
$K_{(3,1)}$ the $\repre{28}$ breaks up as
\begin{equation*}
\repre{28} \to (\repre{1},\repre{1},\repre{3}) \oplus
(\repre{1},\repre{3},\repre{1}) \oplus 2\,
(\repre{3},\repre{1},\repre{1}) \oplus (\repre{1},\repre{2},\repre{2})
\oplus (\repre{3},\repre{2},\repre{2})~,
\end{equation*}
which shows that the $-\half$ eigenvalue is degenerate.  Similar
considerations hold for the $(3,2)$ and $(3,3)$ geometries.  The
generalised self-dual Yang--Mills equations have not been studied for
these geometries.  They may provide interesting an interesting
refinement to the self-dual Yang--Mills equations in quaternionic
K\"ahler geometry.

\section{Conclusion}
\label{sec:fin}

In this paper we have presented a survey of some of the calibrated
geometries which have occurred in recent studies on the local
singularity structure of supersymmetric fivebranes in $\M$-theory
\cite{AFS-cali,AFS-groups,AFSS}.  Some of these geometries appeared
explicitly in \cite{GibbonsPapadopoulos,GLW} and implicitly in some
earlier work
\cite{BerkoozDouglasLeigh,Townsend-Mbranes,Jabbari,OhtaTownsend}.
Calibrated geometries have also appeared in related contexts in other
papers \cite{GT,BBS,BSV,OOY,BBMOOY,Sonia}.  Calibrated geometry is
therefore beginning to emerge as the natural language in which to
phrase geometric questions in the study of branes.  An appropriate
slogan might be
\begin{equation*}
\shadowbox{\em brane geometry is calibrated geometry}
\end{equation*} 

Not all geometries which have appeared in our work have been showcased
here.  Our choice reflects the present level of knowledge in this
topic.  We have omitted two of the subgeometries of $G(5|10)$ which
were obtained in \cite{AFS-groups}, because we were not able to
identify them.  They are summarised in Table~\ref{tab:omitted} below.
The systematic study of the faces of $G(p|n)$ has alas stopped short
of the interesting $G(5|10)$ case: only partial results are known for
$G(4|8)$ and very little indeed for $n>8$.  It is hoped that this
survey might help to rekindle the interest in this problem.

\begin{table}[h!]
\centering
\setlength{\extrarowheight}{5pt}
\begin{tabular}{|*{5}{>{$}c<{$}|}}
\hline
\multicolumn{1}{|c|}{}&
\multicolumn{1}{c|}{Group}&
\multicolumn{1}{c|}{Isotropy}&
\multicolumn{1}{c|}{Geometry}&
\multicolumn{1}{c|}{Fraction}\\
(p|n)&
G&
H&
G/H&
\nu\\[3pt]
\hline\hline
(5|10) & \U_1\times \SU_2 & ? & ? & \tfrac{3}{32}\\
(5|10) &\U_1 & \{1\} & G(1|2)? & \tfrac{1}{8}\\[3pt]
\hline
\end{tabular}
\vspace{8pt}
\caption{Geometries omitted from Table~\ref{tab:summary}.}
\label{tab:omitted}
\end{table}

Finally, it should be mentioned that the calibrated subgeometries of
the grassmannians $G(p|n)$ are far richer than what has been surveyed
in this paper.  We have only looked at geometries defined by {\em
supersymmetric\/} brane configurations; whereas other calibrated
geometries describe non-supersymmetric configurations whose study
might still be physically interesting, since they correspond to local
singularities of minimal submanifolds, which presumably still give
rise to stable states.

\section*{Acknowledgements}

It is a pleasure to thank Bobby Acharya, Bill Spence and Sonia Stanciu
for their collaboration in \cite{AFS-cali,AFS-groups,AFSS} and for
their comments on the present paper.  Thanks are also due to Jerome
Gauntlett for prompting me to look in more detail at the complex
lagrangian geometry, which formed the kernel for Sections
\ref{sec:tour} and \ref{sec:selfdual}.  This work is funded by the
EPSRC under contract GR/K57824 and I would like to thank them for
their support.

\end{document}